\documentstyle[12pt]{article}

\newcommand{\IR}{\hbox{{\rm I}\kern-.2em\hbox{\rm R}}}
\newcommand{\IC}{\hbox{{\rm I}\kern-.6em\hbox{\bf C}}}
\newcommand{\half}{{\scriptstyle\frac{1}{2}}}
\newcommand{\ov}{\overline}
\newcommand{\hf}{\hat{f}}
\begin{document}%

\title{\Large
Realization of compact Lie algebras in K\"{a}hler manifolds}
\author{
{\normalsize\sc D. Bar-Moshe and M. S. Marinov}\\
{\normalsize\it
Department of Physics, Technion-Israel Institute of Technology}\\
{{\normalsize\it
Haifa 32000, Israel}} }

\date{}

\maketitle

\begin{abstract}
The Berezin quantization on a simply connected homogeneous K\"{a}hler
manifold, which is considered as a phase space for a dynamical system,
enables a description of the quantal system in a (finite-dimensional) Hilbert
space of holomorphic functions corresponding to generalized coherent states.
The Lie algebra associated with the manifold symmetry group is given in
terms of first-order differential operators. In the classical theory, the
Lie algebra is represented by the momentum maps which are functions on
the manifold, and the Lie product is the Poisson bracket
given by the K\"{a}hler structure. The K\"{a}hler potentials
are constructed for the manifolds related to all compact semi-simple
Lie groups. The complex coordinates are introduced by means of the Borel
method. The K\"{a}hler structure is obtained explicitly for any unitary group
representation. The cocycle functions for the Lie algebra and the Killing
vector fields on the manifold are also obtained.
\end{abstract}


\newpage
\section{Introduction}

Action of any Lie group on its homogeneous spaces is determined by
realization of the corresponding Lie algebra by means of first-order
differential operators and the Killing vector fields.
Under some conditions, the group representations can be constructed
in terms of linear operators in a Hilbert space of functions on the
manifold. In general, this construction is called {\em geometric
quantization}, and the Hilbert space may be considered as a space of states
of a quantum dynamical system. Moreover, the manifold is the phase space
of a classical dynamical system with an appropriate symplectic structure
and Poisson brackets, representing the Lie algebra of the
transformation group. An important class of such
homogeneous manifolds, provided with a complex K\"{a}hler structure,
has been constructed  on the basis of the Borel theory\cite{borel}.
A general theory of quantization on  K\"{a}hler manifolds was
constructed by Berezin\cite{ber72,ber74,ber75}. In particular,
examples of homogeneous manifolds were given in his works. Another approach
to geometric quantization is based upon the method of coadjoint orbits
proposed by Kirillov\cite{kirill70} and Kostant\cite{kostant}. (Reviews
may be found in refs.\cite{kirill76,hurt,kirill88}.)
The simplest example, where the group is $SU(2)$ and the manifold is $S^2$,
was given previously by Souriau\cite{souriau}.

The purpose of this work is an explicit construction of the Berezin
quantizations for compact Lie groups. Any unitary group representation
corresponds to a (finite, for compact groups) Hilbert space
of holomorphic functions. The system states can be considered also as
{\em generalized coherent states}, introduced by Perelomov\cite{perelomov}.
 Section 2 is a short review of the Berezin quantization, i.e. construction
of the Hilbert space of holomorphic functions on K\"{a}hler manifolds.
Section 3 is an exposition of two realizations of the Lie algebra on
homogeneous K\"{a}hler manifolds; first, in terms of commutators of
first-order differential operators, second, in terms of suitably defined
Poisson brackets
for functions on the manifolds. In Section 4, the Borel parametrization is
introduced for homogeneous manifolds, and the K\"{a}hler potentials are
constructed, using the projection matrices of Ref.\cite{bando}.
Explicit expressions for operator symbols and momentum maps in the local
coordinates are given there, which is the main result of this work.
The subject of Section 5 is a construction of the differential operators
representing the Lie algebra in the space of holomorphic functions on the
manifold. Appendix contains i) the derivation of an explicit
expression for the K\"{a}hler potentials and the momentum maps,
ii) an example: the homogeneous manifolds for unitary groups.

The notations follow standard texts, e.g. the book by Kobayashi and Nomizu
\cite{nomizu}.
The sum over repeated indices is implied throughout the paper.

\section{Berezin quantization on compact \hfill\break
K\"{a}hler manifolds}

Let $\cal M$ be a K\"{a}hler manifold, and $K(z,\bar{z})$ be the K\"{a}hler
potential, defined in any open coordinate neighbourhood of the manifold
with local complex coordinates $z_\alpha$ ($\alpha=1,\cdots,m \equiv$ dim
$\cal M$). If a group $\cal G$ is acting  holomorphically on $\cal M$,
$z\rightarrow gz,\forall g\in {\cal G}$,
the K\"{a}hler potential is transformed according to
    \begin{equation}
K(gz,\overline{gz})=K(z,\bar{z})+\Phi(z;g)+\overline{\Phi(z;g)},
\end{equation}
where $\Phi(z;g)$ is locally a holomorphic function of $z$. This function
must satisfy the cocycle condition,
             \begin{equation}
\Phi(z;g_2g_1)=\Phi(g_1z;g_2)+\Phi(z;g_1),\;\;\;\;
       \forall g_1,g_2\in {\cal G},          \end{equation}
which results from the group property
$z\rightarrow g_2(g_1z)\equiv (g_2g_1)z$.

The corresponding  K\"{a}hler (1,1)-form has the following coordinate
representation
     \begin{equation}
\omega\equiv\omega_{\alpha\bar{\beta}}(z,\bar{z})
dz^\alpha\wedge d\bar{z}^\beta;\;\;\;
\omega_{\alpha\bar{\beta}}=\partial_\alpha\partial_{\bar{\beta}}K,
  \end{equation}
where $\partial_\alpha\equiv\partial/\partial z^\alpha$,
$\partial_{\bar{\beta}}\equiv\partial/\partial \bar{z}^\beta$.
The form $\omega$ is closed,
$\partial\omega=0,$ $\bar{\partial}\omega=0,$
and invariant under the group transformations, as follows from Eq. (1),
and $\bar{\omega}=-\omega$.
The two-form is called {\em integral}, if being integrated over
any two-dimensional cycle in $\cal M$, it gives an integer, times $2\pi i$.
The corresponding K\"{a}hler potential is also called integral.

Any integral K\"{a}hler potential is associated with a holomorphic
line bundle $\cal L$ over $\cal M$. The holomorphic sections of $\cal L$
will be denoted by $\psi$; they represent states of a quantum-mechanical
system and are given by locally holomorphic {\em wave functions} $\psi(z)$.
Following Berezin\cite{ber74}, a Hilbert space structure is assigned to
$\cal L$ by means of the following $\cal G$-invariant scalar product,
     \begin{equation}
(\psi_2,\psi_1)=C\int_{\cal M}\overline{\psi_2(z)}
\psi_1(z)\exp[-K(z,\bar{z})]d\mu(z,\bar{z}).
   \end{equation}
Here the invariant volume element is the $m$-th power of $\omega$,
$d\mu=i\omega\wedge\cdots\wedge i\omega$ ($m$ times), and $C$
is a normalization constant (see Eq. (10) below). (We assume that the
form $\omega$ is {\em non-degenerate}, so the total volume is nonzero.)
Under the action of the group, the wave functions are
subject to transformations, corresponding to (1), i.e.
    \begin{equation}
(\hat{U}(g^{-1})\psi)(z)=\exp [-\Phi(z;g)]\psi(gz),
    \end{equation}
where $\hat{U}(g)$ is the operator representing the group element
$g$ in the Hilbert space, $\hat{U}(g_1)\hat{U}(g_2)=\hat{U}(g_1g_2)$.

Operators in the Hilbert space, which describe observables of
the quantal system, are represented with their symbols, e.g.
$\hat{A}\rightarrow A(z,\bar{\zeta})$, which are mappings
${\cal M\times M}\rightarrow \IC^\times$. The action of the operators
in the Hilbert space is given in terms of their symbols\cite{ber74},
     \begin{equation}
(\hat{A}\psi)(\zeta)=C\int_{\cal M}A(\zeta,\bar{z})\psi(z)
\exp\left[ K(\zeta,\bar{z})-K(z,\bar{z})\right]d\mu(z,\bar{z}).
    \end{equation}
Berezin proved, in particular, that any element of the Hilbert space
is reproduced by the integral with a unit symbol,
     \begin{equation}
\psi(\zeta)\equiv C\int_{\cal M}\psi(z)
\exp\left[ K(\zeta,\bar{z})-K(z,\bar{z})\right]d\mu(z,\bar{z}),
\;\;\;\forall\psi\in{\cal L}.
    \end{equation}
Moreover, the trace of any operator is given by the integral of its symbol,
     \begin{equation}
{\rm tr}(\hat{A})= C\int_{\cal M}A(z,\bar{z})d\mu(z,\bar{z}).
    \end{equation}
For compact manifolds, the trace of the unity operator $\hat{I}$ exists
and equals the total number of states $N$, which is finite in this case,
     \begin{equation}
\hat{I}\rightarrow I(\zeta,\bar{z})\equiv 1,\;\;\;
{\rm tr}(\hat{I})\equiv N=C\int_{\cal M}d\mu(z,\bar{z}).
    \end{equation}
Constant sections of $\cal L$ belong to the Hilbert space for the compact
case, and in order to fix the normalization we assume that
$\psi(z)\equiv 1$ has the unit norm. As a result, the constant $C$ is
expressed in terms of the volume $V$ of the manifold,
     \begin{equation}
C^{-1}=\int_{\cal M}\exp[-K(z,\bar{z})]d\mu(z,\bar{z})=V/N,\;\;\;
V\equiv\int_{\cal M}d\mu(z,\bar{z}).
    \end{equation}
The Hermitean conjugation in the Hilbert space is represented by the
complex conjugation of the symbol and transposition of its arguments,
     \begin{equation}
\hat{A}^\dagger\rightarrow A^*(\zeta,\bar{z})=
\overline{A(z,\bar{\zeta})}.
    \end{equation}
As follows from Eq. (6), the symbol for product of operators
is given by an integral of product of their symbols (the $*$-product)
     \begin{eqnarray}
\hat{A}\hat{B}\rightarrow (A*B)(\zeta,\bar{\eta})\equiv
(N/V)\int_{\cal M}A(\zeta,\bar{z})B(z,\bar{\eta})\times\nonumber\\
\exp\left[ K(\zeta,\bar{z})-K(z,\bar{z})+K(z,\bar{\eta})-
K(\zeta,\bar{\eta})\right]d\mu(z,\bar{z}).
    \end{eqnarray}
Thus the associative algebra of observables for the quantal system is
constructed completely in terms of the operator symbols.
The symbols are functions of two independent variables, holomorphic in
the first of them, and anti-holomorphic in the second one. The "physical
observables" are the reductions to the phase space, where $\zeta=z$.

\section{The symbol representation of\hfill\break
the Lie algebra}

\subsection{Differential operators}

As soon as the transformation law for the wave functions is given by
Eq. (5), the symbols of the operators representing the group elements
are given in terms of the K\"{a}hler potential and the transition
functions of Eq. (1),
     \begin{equation}
(\hat{U}(g)\psi)(\zeta)=\frac{N}{V}\int_{\cal M}
U_g(\zeta,\bar{z})\psi(z)
\exp\left[ K(\zeta,\bar{z})-K(z,\bar{z})\right]d\mu(z,\bar{z}),
    \end{equation}
where
     \begin{eqnarray}
U_g(\zeta,\bar{z})=\exp\left[ K(g^{-1}\zeta,\bar{z})-K(\zeta,\bar{z})
-\Phi(\zeta,g^{-1})\right]\nonumber\\
\equiv\exp\left[
K(\zeta,\overline{gz})-K(\zeta,\bar{z})-\overline{\Phi^(z,g)}\right].
   \end{eqnarray}
The second equality follows immediately from Eq. (1),
it is actually equivalent to the unitarity condition,
$U_{g^{-1}}(\zeta,\bar{z})=\overline{U_g(z,\bar{\zeta})}$.
The symbol of the group element is subject to the following
transformation law,
     \begin{equation}
\hat{U}(h^{-1})\hat{U}(g)\hat{U}(h)\rightarrow
U_{h^{-1}gh}(\zeta,\bar{z})\equiv U_g(h\zeta,\overline{hz}),
\;\;\;\forall h\in{\cal G}.
    \end{equation}

Let us turn to the Lie Algebra $\bf g$ of the group $\cal G$.
 Introducing a basis $\tau_a$ in $\bf g$,
with $a=1,\cdots, n\equiv {\rm dim}\;{\bf g}$, one has a Cartesian
coordinate representation of the group elements and the corresponding
Lie derivatives $D_a$ in the group manifold
       \begin{equation}
g=\exp(-\xi^a\tau_a),\;\;\;\;
D(\xi)_a\equiv -L^b_a(\xi)\partial/\partial\xi^b,
\end{equation}
where $\xi^a$ are the group parameters, and the field $L^b_a$
is given in terms of the adjoit group representation, so that
$g^{-1}D_ag=\tau_a$. The action of the group $\cal G$ in the manifold
determines the holomorphic Killing fields
    \begin{equation}
\nabla_a=\kappa^\alpha_a(z)\partial_\alpha,\;\;\;\;
\kappa^\alpha_a(z)\equiv D(\xi)_a(gz)^\alpha\mid_{g=e}
    \end{equation}
($e$ is the  unit group element). The conjugate Killing field and the
differential operator $\ov {\nabla_a}$ is defined similarly.
Now the differential operators $\hat{T}_a$, which act upon elements
of the Hilbert space $\cal L$ and represent the basis
in $\bf g$, are obtained from Eq. (5),
\begin{equation}
\tau_a\rightarrow \hat{T}_a=\nabla_a-\varphi_a(z),\;\;\;
\varphi_a(z)=D(\xi)_a\Phi(z;g)\mid_{g=e}.
\end{equation}
The Lie multiplication is given by that for the basis elements,
 \begin{equation}
[\tau_a,\tau_b]=f^c_{ab}\tau_c\;\;\rightarrow\;\;
[\hat{T}_a,\hat{T}_b]=f_{ab}^c\hat{T}_c.
\end{equation}
One can introduce a real basis, where the {\em structure constants}
$f_{ab}^c$ are real, and $\hat{T}_a$ are anti-self-adjoint operators
with respect to the scalar product (4). The same relations are satisfied
by the differential operators $D_a$ and $\nabla_a$, while $\varphi_a(z)$
satisfy the linear differential equations
\begin{equation}
\nabla_a\varphi_b-\nabla_b\varphi_a=f^c_{ab}\varphi_c.
\end{equation}
Solution to these equations is given in Section 5.

\subsection{Symbols}

Elements of the Lie algebra are also represented by their symbols,
which can be obtained from symbols of the group elements,
given in Eq. (14),
\begin{eqnarray}
\hat{T}_a\rightarrow
T_a(\zeta,\bar{z})=-\overline{T_a(z,\bar{\zeta})}=
D_aU_g(\zeta,\bar{z})\mid_{g=e}\nonumber\\
=\nabla_aK(\zeta,\bar{z})-\varphi_a(\zeta)\equiv
-\ov{\nabla_a}K(\zeta,\bar{z})+\overline{\varphi_a(z)}.
\end{eqnarray}
(The first equality is true for the real basis, as soon as
$\hat{T}_a=-\hat{T}_a^\dagger$, the last one follows directly from
(1).) Thus the symbols of the Lie algebra
are expressed in terms of the coefficients of the symplectic
one-form generated by the  K\"{a}hler structure
\begin{equation}
T_a(\zeta,\bar{z})=
\kappa^\alpha_a(\zeta)\Lambda_\alpha(\zeta,\bar{z})-\varphi_a(\zeta),
\end{equation}
where $\Lambda_\alpha$ are coefficients of the one-form
\begin{equation}
dK=\Lambda_\alpha d\zeta^\alpha+\overline{\Lambda_\beta}d\bar{z}^\beta.
\end{equation}
Evidently, the symbols satisfy the following relations,
        \begin{equation}
\partial_\alpha T_a(z,\bar{z})=\omega_{\alpha\bar{\beta}}
\overline{\kappa_a^\beta(z)},\;\;\;        
\partial_{\bar{\beta}}T_a(z,\bar{z})=-\omega_{\alpha\bar{\beta}}
\kappa_a^\alpha(z).
         \end{equation}
The functions $T_a(z,\bar{z})$ are called (equivariant)
{\em momentum maps}. It is seen from Eq. (15), that their transformation
law is given by the adjoint group representation,
\begin{equation}
g^{-1}\tau_ag=A_a^b(g)\tau_b\rightarrow
T_a(g\zeta,\overline{gz})=A^b_a(g)T_b(\zeta,\bar{z}),\;\;\;\;
\forall g\in {\cal G}.
\end{equation}
Applying the Lie equations for $\nabla_a$ and eq. (20), one gets
from Eqs. (21) and (24)
       \begin{equation}
-\omega_{\alpha\bar{\beta}}(\kappa_a^\alpha\overline{\kappa_b^\beta}
-\kappa_b^\alpha\overline{\kappa_a^\beta})=
\nabla_aT_b-\nabla_bT_a=f^c_{ab}T_c.
        \end{equation}
This equality is equivalent to the fundamental property of the momentum
maps: they implement a realization of the Lie algebra $\bf g$ in terms of
the Poisson brackets,
\begin{equation}
\{T_a,T_b\}_{\rm P.B.}=f^c_{ab}T_c.
\end{equation}
We have introduced here the Poisson brackets in $\cal M$, which are
determined by a field $\varpi$ dual to the form $\omega$. Namely,
for any two symbols $A(z,\bar{z})$ and
$B(z,\bar{z})$ the definition is
\begin{eqnarray}
\{A,B\}_{\rm P.B.}\equiv\varpi^{\alpha\bar{\beta}}(
\partial_\alpha A\partial_{\bar{\beta}}B-
\partial_\alpha B\partial_{\bar{\beta}}A)=-\{B,A\}_{\rm P.B.}\\
\omega_{\alpha\bar{\beta}}\varpi^{\alpha\bar{\gamma}}=
\delta_{\bar{\beta}}^{\bar{\gamma}},\;\;\;
\omega_{\alpha\bar{\beta}}\varpi^{\gamma\bar{\beta}}=
\delta_\alpha^\gamma.\nonumber
\end{eqnarray}
Wth this definition,
$\overline{\{A,B\}}_{\rm P.B.}=\{\bar{B},\bar{A}\}_{\rm P.B.}$, and
the Jacobi identity for the Poisson brackets results from the fact that
$\omega$ defined in (3) is a closed form.

The equivalence of the commutators (19) and the Poisson brackets (27)
is the reason why the operator -- symbol correspondence is called
{\em quantization}. One should note that the commutator -- Poisson bracket
correspondence holds only for elements of the Lie algebra $\bf g$, but not
necessarily for any pair of observables, which belong to the envelopping
algebra.

The construction of the K\"{a}hler potentials for the homogeneous
manifolds, described in the next section, enables an explicit derivation
of the momentum maps.

\section{The K\"{a}hler structure on compact \hfill\break
homogeneous manifolds}

\subsection{Flag manifolds}

Let $\cal G$ be a compact simple Lie group, and $\cal T$ be its maximal
Abelian subgroup (the maximal torus). The coset space $\cal F=G/T$
(the flag manifold) is provided with a K\"{a}hler structure \cite{borel}.
We shall introduce a local complex parametrization in $\cal F$
by means of the canonical diffeomorphism $\cal G/T\cong G^{\rm c}/P$.
Here $\cal G^{\rm c}$ is the complexification of $\cal G$;
the parabolic subgroup $\cal P$ satisfies the requirements
$\cal P\supset B$, where $\cal B$ is a Borel subgroup of $\cal G$
and $\cal P\cap G=T$.

We shall employ the canonical basis $\{\tau_a\}=\{h_j,e_{\pm\alpha}\}$
in the Lie algebra $\bf g$, where $j=1,\cdots,r\equiv{\rm rank}({\bf g})$
and $\{\alpha\}\in\Delta^+_{\bf g}$ are the positive roots of ${\bf g}$.
In particular, the Lie products of the basis elements are
    \begin{equation}
[h_j,h_k]=0,\;\;\;
[h_j,e_{\alpha}] =(\alpha\cdot{\bf w}_j)e_\alpha.
    \end{equation}
Here the $\{{\bf w}_j\}$ are the fundamental weight vectors which
constitute a system dual to primitive roots $\{\gamma^{(j)}\}$
in the root space, i.e.
     \begin{equation}
2(\gamma^{(j)}\cdot {\bf w}_{j'})/
(\gamma^{(j)}\cdot\gamma^{(j)})=\delta_{jj'}.
    \end{equation}
(Any positive root $\alpha$ is a sum of the primitive roots with
nonnegative  integer coefficients.) For any unitary irreducible group
representation ${\sf R}_{\bf l}$, its highest weight $\bf l$ is given by
a sum of the fundamental weights with nonnegative integer coefficients,
     \begin{equation}
{\bf l}=\sum^r_{j=1}l^j{\bf w}_{j},\;\;\;
l^j=2(\gamma^{(j)}\cdot{\bf l})/(\gamma^{(j)}\cdot\gamma^{(j)}).
    \end{equation}

Given the canonical basis, the Lie algebra $\bf g$ is splitted into
three subalgebras, $\bf g=g^-\oplus t\oplus g^+$, corresponding to three
subsets of the basis elements, $\{e_{-\alpha}\}$, $\{h_j\}$, $\{e_{\alpha}\}$.
Respectively, the Lie algebra $\bf g^+$ generates a nilpotent subgroup
$\cal G^+\subset G^{\rm c}$, and the Lie algebra
${\bf p=g^-\oplus t}^{\rm c}$ generates $\cal P$.
Any element $g\in {\cal G}^{\rm c}$ has a unique Mackey decomposition,
$g=fp$, where $p\in {\cal P}$ and $f\in {\cal G}^+$ (the decomposition
is valid for all $g$, except for a subset of a lower dimensionality).
The complex parameters which can be introduced in $\cal F$
correspond to the positive roots of $\bf g$,
     \begin{equation}
f(z)=\exp(\sum_{\alpha\in\Delta^+_{\bf g}}z^\alpha e_\alpha),
\;\;\;z^\alpha\in\IC.
    \end{equation}
As soon as  $f(z)$ is an element of a nilpotent group,
its matrix representations are polynomials of $z^\alpha$.
 The local form (32) for $f$ is valid in a neighbourhood of the point
$z^\alpha=0$, i.e. the origin of the coordinate system in ${\cal F}$.
Of course, the origin is not a special point, since the
manifold is homogeneous; it is just related to the choice of coordinates.
Transition to other domains of $\cal F$, covering the manifold completely,
can be performed by means of the group transformations.

The group $\cal G^{\rm c}$ acts on $\cal F$ by left multiplications.
Actually, for any element $g$ one has a unique decomposition,
     \begin{equation}
gf(z)=f(gz)p(z;g),\;\;\;\;p(z;g)\in {\cal P},
    \end{equation}
and $gz$ is a {\em rational} function of $z$. For any element $g$
which does not drive the point with coordinates $z^\alpha$ outside
the coordinate neighbourhood containing the origin where (32) is valid,
$gz$ and $p(z;g)$ can be obtained from equation (33) by means of a
linear algebra. Performing two consecutive transformations,
like in Eq. (2), one gets
     \begin{equation}
p(z;g_2g_1)=p(g_1z;g_2)p(z,g_1),\;\;\;\;
  \forall g_1,g_2\in {\cal G}^{\rm c}.        \end{equation}
The Lie equation for the holomorphic  one-cocycles is derived from
this cocycle condition in Section 5.
In view of Eqs. (29), the decomposition (33) shows
that $\cal T$ is a little group of $\cal F$.

\subsection{Fundamental K\"{a}hler potentials}

Solution to equation (1) was found by Bando, Kuratomo,
Maskawa and Uehara\cite{bando} (see also ref.\cite{itoh}).
The general solution is given in terms of projection matrices $\eta_j$,
which exist in any matrix representation of ${\cal G}^{\rm c}$ and
correspond to elements of the Cartan subalgebra $h_j\in{\bf g}$.
Their basic properties are as follows,
     \begin{eqnarray}
\eta_j=\eta_j^\dagger ,\;\;\;\eta_j^2=\eta_j,\;\;\;
\eta_j\hat{h}_k=\hat{h}_k\eta_j,\;\;\;\forall j,k=1,\cdots,r;
\nonumber\\
\eta_j\hat{e}_{-\alpha}\eta_j=\hat{e}_{-\alpha}\eta_j,\;\;\;
\eta_j\hat{e}_{\alpha}\eta_j=\eta_i\hat{e}_{\alpha}.
    \end{eqnarray}
(The hat stands for the matrix representation.) All $\eta_j$ are commuting
with each other. Respectively, in the group representation one has
     \begin{equation}
\eta_j\hat{f}\eta_j=\eta_j\hat{f},\;\;\;\forall f\in{\cal G}^+;\;\;\;\;\;
\eta_j\hat{p}\eta_j=\hat{p}\eta_j,\;\;\;\forall p\in{\cal P}.
    \end{equation}
For any representation of $\cal G^{\rm c}$, where $\hat{h}_j$
are diagonal, all $\eta_j$ are also diagonal. Each $\eta_j$ has $1$ on the
diagonal where $\hat{h}_j$ has its minimum eigen-values;
all the other elements are $0$. As shown in ref.\cite{bmar}, $\eta_j$ can be
expressed in terms of elements of $\cal G^{\rm c}$.
For any given $\bf g$, the matrices $\eta_j$ have different ranks, which are
listed in Appendix B for the simple Lie algebras.

Given $\eta$, the {\em projected determinant} is defined for any matrix $M$,
     \begin{equation}
{\det}_\eta M \equiv\det(\eta M\eta+I-\eta).
   \end{equation}
The operation is designed to be multiplicative for the Mackey decomposition,
as follows from (36). Now the {\em fundamental} K\"{a}hler potentials and
the corresponding transition functions are defined for any $j$, namely
\cite{bando}
     \begin{eqnarray}
K^j(z,\bar{z})\equiv
\log{\det}_{\eta_j}\left(\hat{f}(z)^\dagger\hat{f}(z)\right),\\
\Phi^j(z;g)=-\log{\det}_{\eta_j}\hat{p}(z,g).
    \end{eqnarray}
The transformation property (1) follows immediately from (33) and (36).
The fundamental representation of $\cal G^{\rm c}$ should be used for
this construction. One can show that
the use of the fundamental representation is essential in order
to get {\em all} integral K\"{a}hler two-forms from the
basic potentials (38). As soon as $\hf(z)$ is a polynomial,
$\exp [K^j(\zeta,\bar{z})]$ is also a polynomial in both sets of its arguments.
As we show in Appendix, the fundamental K\"{a}hler potentials can be
represented also in another, sometimes more suitable form,
      \begin{equation}
K^j(\zeta,\bar{z})=
\log{\det}'\left(\hat{f}(\zeta)\eta_j\hat{f}(z)^\dagger\right),
      \end{equation}
where ${\det}'M$ is the notation we use for product of all {\em nonzero}
eigen-values of $M$. (The projection matrices $\eta$ are singular.)
Similarly, the transition functions are given by
      \begin{equation}
\Phi^j(z;g)=-
\log{\det}'\left(\hat{g}\hat{f}(z)\eta_j\hat{f}(gz)^{-1}\right).
      \end{equation}

Now for any given unitary group representation specified with a dominant
weight $\bf l$, Eq. (31), one gets the general expressions for the
corresponding K\"{a}hler potential and the transition function,
  \begin{eqnarray}
K^{\bf (l)}(\zeta,\bar{z})=\sum^r_{j=1}l_jK^j(\zeta,\bar{z}),\\
\Phi^{\bf (l)}(z;g)=\sum^r_{j=1}l_j\Phi^j(z;g).
   \end{eqnarray}
Respectively, $\exp K^{\bf (l)}$ is a polynomial, and its
degree is determined by $\bf l$. The standard orthogonalization procedure
leads to construction of a (finite) polynomial basis $\{\phi_\nu (z)\}$,
starting from $\varphi_1=1$,
    \begin{equation}
\exp [K^{\bf (l)}(\zeta,\bar{z})]=\sum^N_{\nu=1}\phi_\nu(\zeta)
\ov{\phi_\nu(z)}.
    \end{equation}

Having the fundamental potentials, one gets the desired representation of
the Lie algebra.

\subsection{Construction of the momentum maps}

The explicit expressions for the K\"{a}hler potential, Eqs. (38) - (41),
enable one to get a compact form for the symbol of the group element in
Eq. (14). As follows from Eq. (33),
     \begin{equation}
U_g^{\bf (l)}(\zeta,\bar{z})=\\
\prod^r_{j=1}\left[ \frac
{\det_{\eta_j}\left(\hf(z)^\dagger\hat{g}^{-1}\hf(\zeta)\right)}
{\det_{\eta_j}\left(\hf(z)^\dagger\hf(\zeta)\right)}\right]^{l_j}
\equiv\prod^r_{j=1}\left[ \frac
{{\det}'\left(\hat{g}^{-1}\hf(\zeta)\eta_j\hf(z)^\dagger\right)}
{{\det}'\left(\hf(\zeta)\eta_j\hf(z)^\dagger\right)}\right]^{l_j}.
    \end{equation} 
In the local coordinates the symbol of any group element is a rational
function, its numerator being a polynomial with coefficients determined
by the group element, while the denominator can be always chosen as the
reproducing kernel (44).

In order to get the symbols for the basis elements of the Lie algebra
$\bf g$, $\tau_a\rightarrow T^{\bf (l)}_a(\zeta,\bar{z})$,
one has to apply the Lie derivative, Eq. (16). The result is
     \begin{equation}
T_a^{\bf (l)}(\zeta,\bar{z})=
-{\rm tr}\left(\rho^{\bf (l)}(\zeta,\bar{z})\hat{\tau}_a\right).
    \end{equation}
Here the trace is taken in the fundamental representation, and
the symbols depend on $\bf l$ linearly,
     \begin{eqnarray}
\rho^{\bf (l)}(\zeta,\bar{z})=
\sum^r_{j=1}l_j\rho_j(\zeta,\bar{z}),\\
\rho_j(\zeta,\bar{z})\equiv
\hf(\zeta)\eta_j\left(\eta_j\hf(z)^\dagger\hf(\zeta)\eta_j+I-
\eta_j\right)^{-1}\eta_j\hf(z)^\dagger.\nonumber
    \end{eqnarray}
The fundamental projection matrices $\rho_j$ have the following
properties, which can be obtained using Eqs. (45) from
the expression above,
     \begin{eqnarray}
\rho_j(\zeta,\bar{z})^\dagger=\rho_j(z,\bar{\zeta}),\;\;\;\;\;
\rho_j(\zeta,\bar{z})^2=\rho_j(\zeta,\bar{z}),\\
{\rm tr}[\rho_j(\zeta,\bar{z})]={\rm tr}(\eta_j),\;\;\;\;\;
\rho_j(0,0)=\eta_j.
    \end{eqnarray}
Thus $\rho_j(z,\bar{z})$ can be considered as $\eta_j$ transported from the
origin of $\cal M$ to an arbitrary point, as both matrices have the same
set of eigen-values. Sometimes, these matrices are
also called momentum maps\cite{hitchin,odz}, but we
retain this term for their components, given by Eq. (46).
Another explicit expression for $\rho_j$ is derived in Appendix,
     \begin{equation}
\rho_j (\zeta,\bar{z})=
\exp\left[-K^j(\zeta,\bar{z})\right]F_jQ_{F_j}(F_j),\;\;\;
F_j\equiv\hf(\zeta)\eta_j\hf(z)^\dagger,
    \end{equation}
where $Q_F$ is a polynomial defined in Appendix,
and its coefficients are also polynomials
of $(\zeta,\bar{z})$. The degree of $Q_F$ is $[$rank$(\eta)-1]$, and
$Q_{F_j}\equiv I$ if $\eta_j$ has only one unit eigen-value.

As follows from Eq. (33), the $\rho$-matrices are invariant, i.e.
     \begin{equation}
\rho (g\zeta,\ov{gz})=g\rho (\zeta,\bar{z})g^\dagger,
\;\;\;\forall g\in{\cal G},
    \end{equation}
which leads to the equivariance of the momentum maps, Eq. (25).

Thus a group representation is constructed in the space of holomorphic
sections in $\cal L$, which according to the
Borel -- Weil -- Bott theorem is the representation having the highest
weight $\bf l$. If some of the components of $\bf l$ are zero, the little
group is essentially larger than the maximal torus $\cal T$, the form
$\omega$ is degenerate, and neither the invariant volume, nor the
Poisson brackets can be introduced in $\cal F$. For such representations,
the desired  K\"{a}hler manifold is a section of the flag manifold,
$\cal M\subset F$. The contraction is considered in the next subsection.

\subsection{The K\"{a}hler manifolds of lower ranks}

Equations (24) show that $\omega$ is degenerate if there is a number of
pairs $(a,\alpha)$, for which $\partial_\alpha T_a=0$. Let us calculate the
derivative using the representation (46). Applying the Lie derivative (16)
to the definition (46), one gets differential equations for
$\rho(\zeta,\bar{z})$,
     \begin{eqnarray}
D(\zeta)_\alpha\rho_j=\hat{e}_\alpha\rho_j-\rho_j\hat{e}_\alpha\rho_j,\\
\overline{D(z)_\alpha}\rho_j=
\rho_j\hat{e}_\alpha-\rho_j\hat{e}_\alpha\rho_j.\nonumber
    \end{eqnarray}
If $\hat{e}_\alpha$ commutes with $\rho_j$, the r.h.s. vanishes.
At the origin, $z=0$, this is the case if $(\alpha\cdot{\bf w}_j)=0$,
cf. Eq. (29). Therefore if $({\bf l}\cdot\sigma)=0$
for a number of root vectors $\sigma\in\Delta^+_{\bf s}\subset
\Delta^+_{\bf g}$, the K\"{a}hler (1,1)-form is degenerate at the origin,
since the matrix $\omega_{\alpha\bar{\beta}}$ has a number of zero
eigenvalues, as follows from Eq. (24). In this case, the (1,1)-form is
degenerate everywhere in $\cal F$, because it is transported homogeneously
from the origin. The set of the root vectors $\Delta^+_{\bf s}$ orthogonal
to $\bf l$ is generated by a set of primitive roots $\gamma_j$ for which
$l_j=0$. Correspondingly, the Lie algebra $\bf s\subset g$ has the basis
elements $e_\sigma$.

In the other words, if $l_j=0$ for some $j$, the K\"{a}hler structure
on $\cal F$ is degenerate. Now $\cal F$ can be considered as
a fiber bundle, $\cal M$ being its base, where the unitary
group representation ${\sf R}_{\bf l}$ generates a non-degenerate
K\"{a}hler structure. The local coordinates on $\cal M$ are introduced
by restriction $z_\sigma=0$ for $\sigma\in\Delta^+_{\bf s}$. Respectively,
the little group of $\cal M$ is larger than the maximal torus,
     \begin{equation}
\cal M=G/H,\;\;\;\;H=S\otimes T',
     \end{equation}
where $\cal S$ is the semi-simple Lie group having $\bf s$ as its Lie
algebra, and $\cal T'\subset T$ is a torus generated by those basis
elements $h_j$, for which $l_j\neq 0$. This construction has a clear
interpretation in terms of Dynkin graphs\cite{bordem}.
Given a group representation ${\sf R}_{\bf l}$, one has to eliminate from
the Dynkin graph the nodes for which $l_j\neq 0$. The number of such nodes
is $k\leq r$, it may be called the {\em rank} of $\cal M$. The remaining
nodes indicate a semi-simple Lie algebra $\bf s\subset h\subset g$.
As soon as the fundamental K\"{a}hler potentials are constructed for $\cal F$,
the reduction to $\cal M$ is obtained by constraints $z_\sigma=0$ in the
definition of $f(z)$ in Eq. (32). The parabolic subgroup $\cal P$ is
extended respectively; its Lie algebra is
${\bf p=g^-+t'+s}^{\rm c}$.

The representation dimensionality, given by the Weyl formula, can be
represented also in terms of integrals on $\cal M$, see Eq. (10),
     \begin{equation}
N_{\bf l}=\prod_{\alpha\in\Delta^+_{\bf g}}
\frac{(\alpha\cdot ({\bf l}+\rho))}{(\alpha\cdot\rho)}=
\frac{\int_{\cal M}d\mu(z,\bar{z})}
{\int_{\cal M}\exp[-K(z,\bar{z})]d\mu(z,\bar{z})}
     \end{equation}
where $\rho\equiv\sum_j{\bf w}_j$. Actually, the product here is taken for
$\alpha\in\Delta^+_{\bf g}\setminus\Delta^+_{\bf s}$, since the weight vector
$\bf l$ is orthogonal to all roots belonging to $\bf s$.

\section{Holomorphic cocycles and Killing vectors}

Applying the Lie derivative to the element of the parabolic subgroup
$\cal P$, as given in Eq. (33), one transports the basis of the
Lie algebra, which looks like transforming the gauge fields,
       \begin{equation}
\hat{\theta}_a(z)\equiv D_ap(z;g)|_{g=e}
=\hf(z)^{-1}\tau_a\hf(z)-\hf(z)^{-1}\nabla_a\hf(z)\;\;\in{\bf p}.
        \end{equation}
The holomorphic cocycles in Eqs. (18) and (22) are expressed in terms
of the fundamental representation of the Lie algebra,
       \begin{equation}
\varphi_a(z)=\sum_jl_j{\rm tr}\left(\eta_j\hat{\theta}_a(z)\right).
        \end{equation}
The cocycle condition (34) is equivalent to the following set of
differential equations
       \begin{equation}
\nabla_a\hat{\theta}_b(z)-\nabla_b\hat{\theta}_a(z)+
[\hat{\theta}_a(z),\hat{\theta}_b(z)]= f^c_{ab}\hat{\theta}_c(z).
        \end{equation}
Equalities (20) follow immediately. It is noteworthy that the l.h.s.,
like the gauge field strengths, is a curvature tensor on the manifold.

As soon as $\theta_a(z)$ belongs to the Lie algebra $\bf p=g\setminus g^+$,
the following equations hold ($\forall a$ and
$\forall\beta\in\Delta^+_{\bf g}$) which can be used to get explicit
expressions for the Killing fields
       \begin{equation}
{\rm tr}\left(\hat{e}_\beta^\dagger\hat{\theta}_a(z)\right)=0=
{\rm tr}\left(\hat{e}_\beta^\dagger\hf(z)^{-1}{\tau}_a\hf(z)\right)-
\kappa_a^\alpha
{\rm tr}\left(\hat{e}_\beta^\dagger\hf(z)^{-1}\partial_\alpha\hf(z)\right).
        \end{equation}
The Killing fields are obtained now in terms of the adjoint group
representation $A(g)$. Let us use the following notations
(capitals stand for the adjoint representation matrices)
 \begin{equation}
Z\equiv z^\alpha E_\alpha,\;\;\; A(z)=\exp(-Z),\;\;\;
B(z)=(I-A(z))/Z.
          \end{equation}
Here $E_\alpha$ is the adjoint representation of $e_\alpha\in{\bf g^+}$,
so $Z$ is a nilpotent triangular matrix, and the matrices
$A(z)$ and $B(z)$ are polynomials in $Z$. The matrix $B(z)$
appears in the Cartan -- Maurer one-form,
which takes its values in the Lie algebra $\bf g$,
       \begin{equation}
f(z)^{-1} df(z)=dz^\alpha B_\alpha^a(z)\tau_a.
      \end{equation}
Thus the Killing vector fields $\kappa^\alpha_a$ satisfy the following
set of linear equations
       \begin{equation}
\kappa^\alpha_a(z)B^\beta_\alpha(z)= A^\beta_a(z).
      \end{equation}
The solution exists, as soon as the minor of
$B(z)$ corresponding to the Borel subalgebra does not vanish;
$\kappa^\alpha_a(z)$ is a rational function of $z$.

\section{Conclusion}

The results presented here can be summarized as follows.
For any unitary representation ${\sf R}_{\bf l}$ of a compact
simple group $\cal G$, one can construct a compact homogeneous
K\"{a}hler manifold $\cal M\equiv G/H$ and a Hilbert space $\cal L$
of (locally) holomorphic functions which can be considered as a line bundle
upon $\cal M$. The Lie algebra $\bf g$ is realized in $\cal L$ by means of
linear differential operators, or by means of functions on $\cal M$,
called symbols or momentum maps. If ${\sf R}_{\bf l}$ is a nondegenerate
representation, i.e. projections of the dominant weight vector $\bf l$ upon
all primitive roots are positive integers, $\cal H$ is the maximal torus,
$\cal M$ is the flag manifold, and its complex dimensionality equals the
number of the positive roots of $\bf g$. Otherwise, $\cal M$ is a section of
the flag manifold and $\cal H$ is a torus times a simple group.
The K\"{a}hler potential for $\cal M$ is constructed explicitly in terms of
the fundamental group representation. It is given as a superposition of
fundamental potentials with positive integer coefficients, Eq. (42).

The manifold $\cal M$ can be considered as a phase space of a
dynamical system, and elements of the Hilbert space represent
the generalized coherent states. The Poisson brackets
are derived from the K\"{a}hler structure, providing a realization
of the Lie algebra $\bf g$ in terms of the momentum maps. From this
point of view, the construction described here is the Berezin
quantization on homogeneous manifolds.

The K\"{a}hler structure
can be introduced in $\cal M$ by means of Eq. (42) with arbitrary
coefficients $l_j$, and one gets a representation of the Lie algebra.
Then the manifold can be still considered as a phase space of a dynamical
system, but quantization (and the group representation) is possible only
for integer coefficients.

Extension of the present approach to non-compact groups and
infinite-dimensional (universal) groups is the subject of a
future work.

The support to this research from G. I. F. and the Technion V. P. R.
Fund is gratefully acknowledged.

\section*{Appendix}

\subsection*{A. Derivation of Eqs. (40) and (50)}

We shall consider the fundamental K\"{a}hler potential, given in Eq. (38)
according to ref.\cite{bando}, as a limit at $t\rightarrow 1$ of the
following function
     \begin{equation}
K_t=\log\det [I-t\eta(I-M)\eta]\equiv
{\rm tr}\log[I-t\eta(I-M)\eta],\;\;\;M\equiv\hf^\dagger\hf.
     \end{equation}
Using the expansion in powers of $t$ at $t<1$, one gets for $\eta=\eta^2$
\begin{equation}
K_t={\rm tr}\left\{\log [I-t(I-F)]-(I-\eta)\log (1-t)\right\},\;\;\;
F\equiv\hf\eta\hf^\dagger.
\end{equation}
Let us introduce the following notations: $N$ -- the dimensionality of
the matrix representation, $P_F(\lambda)\equiv\det(F-\lambda I)$ -- the
characteristic polynomial for matrix $F$,
$\nu={\rm tr}(I-\eta)$ -- the number of zero
eigen-values of $\eta$ (and of $F$), $\tau=(1-t)/t$. Now the result in
Eq. (40) is evident from
\begin{eqnarray}
K_t=\log\frac{P_F(-\tau)}{\tau^\nu}-(N-\nu)\log (1+\tau),\;\;\;\\
K\equiv\lim_{\tau\rightarrow 0}K_t=\log{\det}'F.\nonumber
\end{eqnarray}
(The function $\log\det '$, where $\det '$ is the Fredholm determinant
of an elliptic operator in the Hilbert space with its zero modes
excluded, appears also as the derivative of the operator $\zeta$-function
at zero, cf. e.g.\cite{schwarz}.)

Similarly, one gets Eq. (49) for the matrix $\rho$, which can be
represented by the same limit, namely
      \begin{equation}
\rho=\lim_{\tau\rightarrow 0}F(F+\tau I)^{-1}.
      \end{equation}
In the other words,
the eigen-values of $\rho$ vanish together with the eigen-values
of $F$, and are equal to $1$ otherwise.
As soon as the rank of $F$ equals the rank of $\eta$, one can write the
characteristic polynomial as follows,
      \begin{equation}
P_F(\lambda)=(-\lambda)^\nu[D_F-\lambda Q_F(\lambda)],\;\;\;
D_F\equiv{\det}'F.
      \end{equation}
Here $Q_F(\lambda)$ is a polynomial of degree $N_1=N-\nu-1$, and
its coefficients are
polynomials in $(\zeta,\bar{z})$, which are expressed in terms of the traces
$f_n={\rm tr }(F^n)$ for $n=1,\cdots,N_1$, namely
      \begin{equation}
Q_F(\lambda)=(-\lambda)^{N_1}+f_1(-\lambda)^{N_1-1}+
\half(f^2_1-f_2)(-\lambda)^{N_1-2}+\cdots\;.
      \end{equation}
The result in Eq. (50) follows, as soon as $D_F=\exp(K)$, which is also a
polynomial in $f_n$.

\subsection*{B. Example: \hfill\break
Unitary groups and the Grassmann manifolds}

Fot ${\cal G}=SU(N)$, the fundamental group representation is
$N$-dimensional, and the index in Eq. (42) is in the interval
$1\leq j\leq r=N-1$. Up to a normalization factor,
the eigenvalues of $\hat{h}_j$ are $-1$ ($j$ times), $j$, and $0$
( $(N-j-1)$ times). The corresponding projection matrix $\eta_j$
has $j$ values $1$, other $(N-j)$ values are $0$. The local coordinates
corresponding to the positive roots are elements of a triangular matrix
$\hat{z}$, i.e. $z_{jk}$, $1\leq j < k\leq N$ (other elements of
the matrix are zero), and the complex
dimensionality of $\cal F$ is $\frac{1}{2}N(N-1)$. The matrix $\hf (z)$
is triangular; its diagonal elements are $1$, and polynomials in $z^\alpha$
stand above the diagonal. The manifold $\cal F$ has an additional symmetry
under a reflection of the root space, so that $j\rightarrow N-j+1$,
and $\eta_{N-j+1}\rightarrow \sigma(I-\eta_j)\sigma^{-1}$, where $\sigma$
is a matrix reversing the order of components. Matrices
$F_j=\hf\eta_j\hf^\dagger$ can be easily obtained in a general form.

The Grassmann manifold ${\cal M}={\sf Gr}(p,q)\equiv
U(p+q)/U(p)\otimes U(q)$ (where $1\leq q\leq p<N\equiv p+q$) is a
rank-one section of $\cal F$. The group representation realized in $\cal M$
is specified with a positive integer $l$, and $l_j=l\delta_{jq}$.
The complex dimensionality of the manifold is $pq$, and the local
coordinates are elements of a $p\times q$ matrix $\hat{z}$, so that the
elements of $\hf$ are $f_{jk}=\delta_{jk}+z_{j,k-p}$, where
$1\leq j\leq p$, and $p+1\leq k\leq p+q$.
The resulting K\"{a}hler potential is
        \begin{equation}
K(z,\bar{z})=l\log\det(I_q+\hat{z}^\dagger\hat{z}).
         \end{equation}
For $q=1$, $\hat{z}$ is a complex vector, ${\cal M}\equiv{\sf CP}^{N-1}$
is the complex projective space, and the Fubini -- Study form\cite{nomizu}
appears from the K\"{a}hler potential.
The metrics is ``quantized", since $l$ is integer.

It is noteworthy that for any compact Lie group the K\"{a}hler structure
can be obtained by restriction from a unitary group, as soon as
the group is embedded in it, ${\cal G}\subset SU(N)$.

\end{document}